\documentclass{elsart}
\usepackage{epsfig}

\def\myDate{(May 7, 2001)}
\def\mySpecialText{DRAFT version 3.9 \myDate}

\def\myspecial#1{}                   

\newcommand{\myfigure}[3]{%
  \begin{figure}[b]
  \begin{center}
    \epsfxsize #1\textwidth \epsfbox{#2.eps}
    \caption{#3}
	\label{fig:#2}
  \end{center}\end{figure}}

\def\Fig#1{Fig.~\ref{fig:#1}}

\def\myBR{3.36}              
\def\myBRcomm#1{\mbox{(#1)}}
\def\myBRstat{\pm0.53\myBRcomm{stat}}
\def\myBRsyst{\pm0.42\myBRcomm{sys}}
\def\myBRtheo{{}^{+0.50}_{-0.54}\myBRcomm{th}}
\def\myBRe{(\myBR\myBRstat\myBRsyst\myBRtheo)\times10^{-4}}

\def\myNcand{222}            
\def\myNexcess{106.5\pm16.8} 
\def\myNbg{115.5\pm7.7}      
\def\myNBBbg{9.1\pm1.8}      

\def\myMBreso{4.4}           
\def\mySNratio{0.9}          
\def\mySFWinpurity{(0.7\pm 0.2)\%}  
\def\mySFWNinpurity{0.7\pm 0.2}     

\def\myEffi{(2.58\pm0.29)\%} 

\def\myEffiLomb{15\%}        
\def\myEffiHimb{16\%}        
\def\myErrRmix{5.0\%}        

\def\myKSeffi{(65\pm4)\%}    
\def\myPZeffi{(60\pm4)\%}    

\def\myPHOeffi{(80\pm4)\%}   
\def\myKAONeffi{65\%}        
\def\myPIONfake{2\%}         
\def\myPreseleffi{(12.8\pm1.3)\%}    
\def\myMXseffi{(57\pm2)\%}    
\def\myMBeffi{(70\pm1)\%}     
\def\mySFWeffi{(50\pm3)\%}    

\def\myKstarBF{(3.8\pm0.9)\times10^{-5}} 

\def\epem{e^+e^-}
\def\btosgam{b{\to}s\gamma}
\def\btoc{b{\to}c}
\def\qqbar{q\bar{q}}

\def\BBbar{B\bar{B}}

\def\PZ{\pi^0}
\def\KS{K^0_S}
\def\Xs{X_s}

\def\BtoXsgam{B{\to}X_s\gamma}
\def\BtoKstgam{B{\to}K^*(892)\gamma}
\def\PZtoGG{\PZ\to\gamma\gamma}
\def\KStoPP{\KS\to\pi^+\pi^-}
\def\BtoDrho{B{\to}D^{(*)}\rho}
\def\etatogg{\eta{\to}\gamma\gamma}
\def\etatopppz{\eta{\to}\pi^+\pi^-\pi^0}
\def\phitokk{\phi{\to}K^+K^-}
\def\Dstophipi{D_s{\to}\phi\pi}
\def\eeg{e^+e^-\gamma}
\def\eetoeeg{e^+e^-{\to}\eeg}
\def\BtoDpi{B^-{\to}D^0\pi^-}
\def\DtoKpi{D^0{\to}K^-\pi^+}

\def\fbinv{\mbox{~fb}^{-1}}
\def\GeVcc{\mbox{~GeV}/c^2}
\def\GeVc{\mbox{~GeV}/c}
\def\GeV{\mbox{~GeV}}
\def\MeV{\mbox{~MeV}}
\def\MeVcc{\mbox{~MeV}/c^2}

\def\calF{F}
\def\BR{{\it Br}}

\def\Ebeam{E_{beam}}
\def\DE{\Delta{E}}
\def\Egam{E_\gamma}
\def\EXs{E_{X_s}}
\def\MB{M_{bc}}
\def\MXs{M_{X_s}}
\def\pgam{\vec{p}_\gamma}
\def\pXs{\vec{p}_{X_s}}
\def\thetags{\theta_{X_s\gamma}}

\begin{document}

\begin{frontmatter}

\hbox to \textwidth{
\lower 2.5cm
\hbox to 2.5cm{
\hss
\epsfysize3cm
\epsfbox{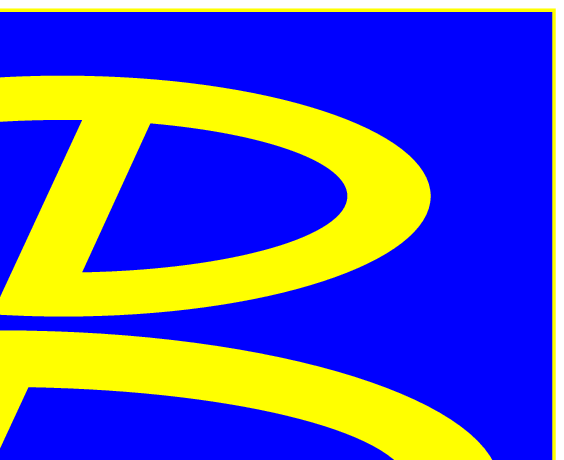}    
}
\hss
\hbox to 3cm{
\begin{tabular}{r}
{KEK preprint 2001-3} \\
{Belle preprint 2001-2}
\end{tabular}
\hss
}
}
\vspace{12pt}

\myspecial{!userdict begin /bop-hook{gsave 300 50 translate 5 rotate
    /Times-Roman findfont 18 scalefont setfont
    0 0 moveto 0.70 setgray
    (\mySpecialText)
    show grestore}def end}

\title{
A Measurement of the Branching Fraction
for the Inclusive $\BtoXsgam$ Decays with the Belle Detector}


\author{The Belle Collaboration} 

\begin{center}
{\normalsize
K.~Abe$^{10}$, 
K.~Abe$^{37}$, 
I.~Adachi$^{10}$, 
Byoung~Sup~Ahn$^{15}$, 
H.~Aihara$^{38}$, 
M.~Akatsu$^{20}$, 
G.~Alimonti$^{9}$, 
K.~Aoki$^{10}$,
K.~Asai$^{21}$, 
Y.~Asano$^{43}$
% ...
}\end{center}

\address{
% ...
$^{9}${University of Hawaii, Honolulu HI}\\
$^{10}${High Energy Accelerator Research Organization (KEK), Tsukuba}\\
% ...
$^{15}${Korea University, Seoul}\\
% ...
$^{20}${Nagoya University, Nagoya}\\
$^{21}${Nara Women's University, Nara}\\
% ...
$^{37}${Tohoku University, Sendai}\\
$^{38}${University of Tokyo, Tokyo}\\
% ...
$^{43}${University of Tsukuba, Tsukuba}\\
% ...
}
\normalsize


\begin{abstract}
We have measured the branching fraction of the inclusive radiative $B$
meson decay $\BtoXsgam$ to be
\[
\BR(\BtoXsgam)=\myBRe.  
\]
The result is based on a sample of $6.07\times10^6$ $\BBbar$ events
collected at the $\Upsilon(4S)$ resonance with the Belle detector at the
KEKB asymmetric $e^+e^-$ storage ring.

\vspace{3\parskip}
\noindent{\it PACS:} 13.40.Hq, 14.40.Nd, 14.65.Fy
\end{abstract}


\end{frontmatter}
\clearpage



In the framework of the Standard Model (SM), the inclusive decay
$\BtoXsgam$, where $X_s$ is a hadronic recoil system containing an $s$
quark, proceeds primarily through the electroweak penguin diagrams of
the radiative $\btosgam$ transition.  These diagrams are of interest
because they have loops where non-SM particles, such as charged Higgs
bosons or SUSY particles, can contribute and potentially produce
deviations from SM expectations.  An SM branching fraction calculation
for $\BtoXsgam$ that includes next-to-leading order QCD corrections has
an expected precision of 10\% \cite{bib:xsgam-sm}.  Experimental
measurements of the branching fraction at this level of precision are
useful for identifying or limiting non-SM theories
\cite{bib:xsgam-nonsm}.

A semi-inclusive analysis is used to reconstruct the $\BtoXsgam$ decay
from a primary photon, a kaon and multiple pions.  The monochromatic
photon energy from the two-body decay $\btosgam$ is smeared by gluon
emission, the Fermi motion of the $b$ quark inside the $B$ meson, and
the $B$ meson momentum in the $\Upsilon(4S)$ center-of-mass (CM) frame.

The $\BtoXsgam$ decay is studied with the Belle detector at the KEKB
asymmetric $\epem$ storage ring \cite{bib:kekb}.  The dataset consists
of a sample of $6.07\times10^6$ $\BBbar$ events corresponding to an
integrated luminosity of $5.8\fbinv$ taken at the $\Upsilon(4S)$
resonance ($\sqrt{s}=10.58\GeV$), and an off-resonance background sample
of $0.6\fbinv$ taken $60\MeV$ below $\Upsilon(4S)$.  The beam energies
at the $\Upsilon(4S)$ resonance are $3.5\GeV$ for positrons and
$8.0\GeV$ for electrons.  Quantities used in this analysis are defined
in the CM frame, in which the beam energy $\Ebeam$ is defined as
$\sqrt{s}/2$.  Inclusion of charge conjugated modes are implied
throughout the text.

A full description of the Belle detector is given in \cite{bib:belle};
here we briefly describe the components relevant for this analysis.
Charged tracks are reconstructed inside a 1.5~T magnetic field induced
by a super-conducting solenoid magnet, by means of a 50 layer central
drift chamber (CDC) that covers the laboratory polar angle
$17^\circ\!<\!\theta_{lab}\!<\!150^\circ$.  Tracks are fitted through
the CDC and a three layer double sided silicon vertex detector (SVD),
with a transverse momentum resolution of
$(\sigma_{p_t}/p_t)^2=(0.0019p_t)^2+(0.0034)^2$ ($p_t$ in
$\mbox{GeV}/c$).  For particle identification, we combine information
from three detectors.  The high momentum range, typically from 1 to
$3.5\GeVc$, is covered by a silica aerogel Cherenkov counter (ACC)
system, providing threshold-type kaon-to-pion separation.  The momentum
range below $1.5\GeVc$ is covered by a time-of-flight (TOF) counter
system, and the very low and high momentum ranges are accessible with
$dE/dx$ information from the CDC.  Photons are detected by an
electromagnetic calorimeter (ECL) located between the particle
identification devices and the solenoid coil.  The entire tracking
acceptance is covered with 8736 CsI(Tl) crystals that are typically
$5.5\times5.5\mbox{~cm}^2$ in cross-section at the front surface and
16.2 $X_0$ in depth.  The photon energy resolution is $(\sigma_E/E)^2 =
{0.013^2+(0.0007/E)^2+(0.008/E^{1/4})^2}$ ($E$ in GeV), which is better
than 2\% for the primary photons used in this analysis.  The Belle
detector is modelled with the GEANT program \cite{bib:geant} to simulate
the detector response.  Both the data and the Monte Carlo (MC) events
\cite{bib:mc} are reconstructed with the same program.


Hadronic events are selected with an efficiency of 99\% for $\BBbar$
final states.  Events with at least three tracks from the interaction
region and visible energy greater than $\!0.2\sqrt{s}$ are selected.
The QED and $\tau^+\tau^-$ events that remain are removed using ECL
information and event topology \cite{bib:hadronB}.

Photon candidates are selected from ECL clusters of $5\times5$ crystals.
Each photon candidate is required to have a laboratory energy greater
than $20\MeV$ with no associated charged track, and shower shape
consistent with an electromagnetic shower.
We accept events in which the most energetic photon is in the barrel
region ($33^\circ\!<\!\theta_{lab}\!<\!132^\circ$).  We require that
95\% of its energy is concentrated in the central $3\times3$ crystals.
The efficiency of these criteria for the primary photon is $\myPHOeffi$.
In order to reject $\PZ$ and $\eta$ mesons, we combine this primary
photon candidate with any other photon (for $\eta$ we only use photons
with $\Egam>200\MeV$) and reject the event if the invariant mass of the
two photons is within $\pm18\MeVcc$ of $M_{\pi^0}$ or $\pm32\MeVcc$ of
$M_\eta$.

We reconstruct the recoil system $\Xs$ in 16 different final states of
one charged kaon or $\KS$ plus one to four pions which may include one
$\PZ$.  We combine the primary photon with every $\Xs$ combination and
calculate the opening angle $\thetags$ between $\pgam$ and $\pXs$, the
energy difference $\DE=\Egam+\EXs-\Ebeam$ and the beam constrained mass
$\MB=\sqrt{\Ebeam^2-|\pgam + \pXs|^2}$.  Then, we require
$\thetags>167^\circ$, $\MB>5.2\GeVcc$ and $-0.15<\DE/\mbox{GeV}<0.1$.

Charged kaon candidates are selected using a kaon-to-pion likelihood
ratio.  For every charged track, likelihoods from ACC, TOF and $dE/dx$
are calculated for pion and kaon hypotheses.  A combined likelihood is
constructed for each hypothesis, and a tight cut is applied on the
likelihood ratio.  Kaons are identified with a typical efficiency of
$\myKAONeffi$ with $\myPIONfake$ pion fake rate.  All charged tracks
that are not identified as kaons are considered to be pions.

The neutral kaon candidates are reconstructed in the $\KStoPP$ mode from
two oppositely charged tracks.  The $\KS$ momentum is recalculated with
a vertex constrained fit.  The candidates are then required to have an
invariant mass within $\pm8\MeVcc$ $(\sim\!2\sigma)$ of $M_{K^0}$ and a
vertex that is displaced from the interaction point in a direction
consistent with the $\KS$ momentum.  The $\KS$ reconstruction efficiency
is $\myKSeffi$.

The $\PZtoGG$ candidates are reconstructed from two photons that have an
invariant mass within $\pm16\MeVcc$ $(\sim\!3\sigma)$ of $M_{\PZ}$.
Then, the $\PZ$ momentum is recalculated with a mass constrained fit.
The $\PZ$ reconstruction efficiency is $\myPZeffi$.

When multiple candidates are found in an event, the best candidate is
selected as follows: If there is at least one charged kaon or pion that
forms the $\Xs$ vertex with the constraint from the run-by-run
determined profile of the $B$ meson decay vertices
\cite{bib:belle-mixing}, the candidate with the largest vertex
confidence level is chosen.  The candidate with the largest $\thetags$
is chosen in the following cases: (1) there is no charged kaon or pion
to form the vertex; (2) there remains an ambiguity whether to add a
$\PZ$; or (3) there are several $\PZ$ and $\KS$ candidates to choose
from.  After selecting the best candidate, we require $\MXs<2.05\GeVcc$.


The largest background source is from continuum $\qqbar$ production
where the photon originates from initial state radiation or from high
momentum neutral hadron decays such as $\PZtoGG$ with one undetected
photon.  In order to reduce and estimate the contribution of the
$\qqbar$ background, we introduce a new variable that exploits the
topological difference between the jet-like $\qqbar$ events and the
spherical $\BBbar$ events.  We define the following variables in the $B$
meson rest frame,
\begin{displaymath}
  R_l = {\sum_i |p_i||p_\gamma|P_l(\cos\theta_{i\gamma}) \over
               \sum_i |p_i||p_\gamma|}
{,~~}
  r_l = {\sum_{i,j} |p_i||p_j|P_l(\cos\theta_{ij}) \over
               \sum_{i,j} |p_i||p_j|},
\end{displaymath}
where $P_l$ is a Legendre polynomial of order $l$, and $i$, $j$ run over
all the photons and charged tracks that are not used to form the $B$
candidate.  Then, we combine six of them into a Fisher discriminant
\begin{displaymath}
  {\calF} = \sum_{l=2,4} \alpha_l R_l
           + \sum_{l=1,2,3,4} \beta_l r_l,
\end{displaymath}
where the six coefficients $\alpha_l$ and $\beta_l$ are optimized to
maximize the discrimination between signal and background as shown in
\Fig{sfw}.  The terms $R_1$ and $R_3$ are excluded from the Fisher
discriminant, because either of these terms is found to have a
correlation with $\MB$.  We call ${\calF}$ the {\em Super Fox-Wolfram}
(SFW) variable, since the terms are combined in such a way as to enhance
the discriminating power of the original Fox-Wolfram moments
\cite{bib:fox-wolfram}.  We select events with ${\calF}>0.1$, which is a
compromise between the statistical significance of the signal and the
size of the systematic error due to background subtraction.  In order to
estimate the $\qqbar$ background contribution, we use the SFW sideband
region of ${\calF}<-1.5$, where the signal fraction is only
$\mySFWinpurity$.  This translates into a contribution of
$\mySFWNinpurity$ events in the signal region.

In addition to the $\qqbar$ background, there is a contribution from
$\btoc$ decays, in which $\BtoDrho$ modes are dominant.  This
background, which does not contribute to the SFW sideband, is estimated
by MC to be $\myNBBbg$ events and is subtracted from the data sample.

The signal yield is obtained from the $\MB$ distribution (\Fig{mb}).  In
\Fig{mbcorr} we show that the ratio between the number of events in the
signal region and in the sideband of the SFW variable is independent of
the beam constrained mass $\MB$ according to a MC calculation assuming
continuum $\qqbar$ production.  This observation is consistent with the
background estimated using the off-resonance data.  Thus, the SFW
sideband data can be used to model the background shape of the $\MB$
spectrum in the absence of a substantial sample of off-resonance data.
We fit the $\MB$ distribution with a background shape taken from the SFW
sideband data and a signal shape obtained from MC simulation.

We observe $\myNcand$ events in the signal region $\MB>5.27\GeVcc$, of
which $\myNbg$ is the estimated background contribution, and is
consistent with the MC expectation.  An excess of $\myNexcess$ events is
clearly seen in the $\MB$ distribution shown in \Fig{mb} at the $B$
meson mass with a mass resolution of $\myMBreso\MeVcc$ and a
signal-to-background ratio of {\mySNratio}.

To obtain the background subtracted recoil mass spectrum shown in
\Fig{mxs}, we use the shape of the $\MXs$ spectrum determined from MC,
whose contribution is normalized to the estimated background of the
$\MB$ distribution.  We adopt this method because the ratio between the
number of signal events and SFW sideband events has a finite slope as a
function of $\MXs$, especially in the region $\MXs>2.05\GeV$, and thus
the SFW sideband does not properly represent the background sample for
the $\MXs$ spectrum \cite{bib:ushi-thesis}.  Nevertheless, the slope in
the $\MXs<2.05\GeV$ region is small and compatible with a flat
distribution within the statistical fluctuation, and the background
subtraction method using the SFW sideband sample leads to consistent
results for the $\MXs$ spectrum and the $\BtoXsgam$ branching fraction.


The signal MC sample is generated as a mixture of the exclusive
$\BtoKstgam$ mode and an inclusive $\BtoXsgam$ contribution for
$\MXs>1.15\GeVcc$ in order to separate the $K^*(892)$ signal from higher
resonances starting with $K_1(1270)$.  The recoil system $X_s$ is
modelled as an equal mixture of $\bar{s}d$ and $\bar{s}u$ states
\cite{bib:mc}.  To simulate the inclusive $\MXs$ spectrum, we adopt the
model by Kagan and Neubert \cite{bib:kagan-neubert} with the $b$ quark
pole mass parameter $m_b=4.75\GeVcc$.

The fraction of $K^*(892)\gamma$ decays in the $\BtoXsgam$ transition,
$r_{K^*\gamma}$, is determined from the $\MXs$ spectrum with a
consideration of the migration effects on the reconstructed $\MXs$ due
to incorrect particle assignments to the $\Xs$ final state.  The $\MXs$
spectrum is divided into two regions: the $K^*(892)$ region below
$1.15\GeVcc$ and the continuum region between $1.15\GeVcc$ and
$2.05\GeVcc$.  The MC sample is analyzed to determine the number of
events migrating between the regions.  The MC results are used to unfold
the migration effects and to generate a set with an improved value,
$r_{K^*\gamma} = (11.3\pm3.5)\%$.  Using the new MC sample, the
procedures of $M_{bc}$ fitting, $\MXs$ background subtraction, and
$r_{K^*\gamma}$ unfolding are iterated.  It is checked that
$r_{K^*\gamma}$ is stable under the iteration.
The size of the migration effects and the value of $r_{K^*\gamma}$
depend on the assumed $m_b$ value: for $m_b=4.75\GeVcc$, 77\% of the
events reconstructed below $2.05\GeVcc$ are genuine and 23\% are from
the region above $2.05\GeVcc$.  We note that due to a non-negligible
amount of migration between the regions below and above $2.05\GeVcc$,
this boundary does not correspond to a sharp value in $\MXs$ or $\Egam$.

According to the authors of ref.~\cite{bib:kagan-neubert}, the
$K^*(892)$ contribution can be approximated by integrating the $\MXs$
spectrum up to a threshold mass above the $K^*(892)$ mass, though their
model does not explicitly take resonance contributions into
account. This integral is sensitive to the parameter $m_b$.  To match
the measured $K^*(892)$ fraction we varied $m_b$ and the threshold mass
and observed that $m_b\geq4.9\GeVcc$ is disfavored.  We use
$m_b=(4.75\pm0.10)\GeVcc$ as the parameter region to test the model
dependence of the branching fraction measurement \cite{bib:ali-priv}.
We assume a somewhat larger error on $m_b$ than that used for the
semileptonic $B$ meson decays in the combined LEP, SLD and CDF
evaluation \cite{bib:lepsldcdf-semilep} based on the recent theoretical
work of \cite{bib:uraltsev}, in which $m_b(1\GeV)=(4.58\pm0.06)\GeVcc$
is quoted.  We note that the corresponding pole mass is
$m_b\simeq4.70\GeVcc$.  Other theoretical input parameters have little
impact on the shape of the $\MXs$ spectrum.


The event selection efficiency is determined from a MC sample that is
calibrated with high statistics control data samples for all final state
particles.  The reconstruction efficiency for the primary photon is
obtained using photons with energies between 2 and 3 GeV from the
processes $\eetoeeg$ and $\etatogg$.  The charged tracking efficiency
for momenta relevant to this analysis is obtained by comparing the high
momentum $\eta$ yields for the $\etatopppz$ and $\etatogg$ decay modes.
The kaon selection efficiency is determined from a sample of $\phitokk$
and $\Dstophipi$ decays.  The pion rejection efficiency is also
determined from the $\etatopppz$ sample.  The $\KS$ and $\PZ$
efficiencies are tested with a sample of $K^*(892)$ decays into
$K^+\pi^-$, $K^+\pi^0$, $K^0_S\pi^+$ and $K^0_S\pi^0$.  Combining all 16
different channels, the selection efficiency becomes $\myPreseleffi$.
The $\MXs$, SFW and $\MB$ cuts reduce the efficiency by $\myMXseffi$,
$\mySFWeffi$ and $\myMBeffi$, respectively.
The SFW cut and the $\pi^0/\eta$ veto efficiencies are measured by
applying the cuts on a sample of exclusively reconstructed $\BtoDpi$,
$\DtoKpi$ decays, in which the first pion is assumed to be massless in
order to mimic the primary photon for the cuts.
The size of the systematic error from the efficiency determination is
dominated by the statistical error of the control sample.  The final
event reconstruction efficiency is $\myEffi$.

The total systematic error is quoted as a quadratic sum of the
systematic errors on the efficiency, the best candidate selection
procedure, signal MC statistics, SFW sideband uniformity and the number
of $\BBbar$ events.  The sensitivity of the cuts to the experimental
error on the beam constrained mass $\MB$ is tested by changing the MC
resolution function, and the effect is found to be negligible.  The
efficiency errors on the 16 different final states are calculated
individually and are then combined.  The best candidate selection is
tested with a $\BtoDpi$ sample, which is analyzed using the inclusive
reconstruction method of this analysis.  The effect of the SFW sideband
uniformity is tested by adding a slope to the background distribution
within the tolerance level of \Fig{mbcorr}.  The contributions to the
systematic error on the branching fraction are given in
Table~\ref{tbl:syserror}.


The fitted signal yield is extrapolated over the entire $\MXs$ region
using the model of ref.~\cite{bib:kagan-neubert} with $m_b=4.75\GeVcc$
and corrected for the final states that are not reconstructed.  The
analysis is repeated for $m_b=4.65\GeVcc$ and $4.85\GeVcc$ to test the
model uncertainty.  The signal efficiency is found to be $\myEffiLomb$
lower and $\myEffiHimb$ higher, respectively.  This error and the error
of $r_{K^*\gamma}$, which typically corresponds to $\myErrRmix$ in the
branching fraction calculation, are combined into the theoretical model
error.

Finally, we obtain the branching fraction of $\BtoXsgam$,
\begin{displaymath}
\BR(\BtoXsgam) = \myBRe
\end{displaymath}
where the first error is statistical, the second is systematic and the
third is the theoretical model error.  Using the values of the measured
$\BtoXsgam$ branching fraction and the fraction of the $K^*(892)\gamma$,
we find the $B\to K^*(892)\gamma$ branching fraction to be $\myKstarBF$
(statistical error only), which is consistent with the $B\to
K^*(892)\gamma$ branching fraction measurements from CLEO
\cite{bib:excl-radb} and a separate Belle analysis
\cite{bib:excl-radb-belle}.  In \Fig{egamma} we show the measured photon
spectrum in which the background is subtracted and the efficiency loss
due to the cut on $\MXs$ is corrected by using MC input.  This spectrum
may be compared with model calculations.


To summarize, we have measured the inclusive branching fraction of
$\BtoXsgam$ decay with the Belle detector.  The result is consistent
with the SM prediction of $\BR(\BtoXsgam)=(3.28\pm0.33)\times10^{-4}$
\cite{bib:xsgam-sm}.  The result is also consistent with previous
measurements by the CLEO \cite{bib:cleo-95} and ALEPH
\cite{bib:aleph-98} experiments.


We gratefully acknowledge the efforts of the KEKB group in providing
us with excellent luminosity and running conditions and the
help with our computing and network systems provided by members
of the KEK computing research center.
We wish to thank M.~Neubert and A.~Ali for helpful suggestions on the
modelling of the $\BtoXsgam$ MC.
We thank the staffs of KEK and collaborating institutions for their
contributions to this work,
and acknowledge support from the Ministry of Education, Science, Sports and
Culture of Japan and
the Japan Society for the Promotion of Science;
the Australian Research Council and the Australian Department of Industry,
Science and Resources;
the Department of Science and Technology of India;
the BK21 program of the Ministry of Education of Korea and
the Basic Science program of the Korea Science and Engineering Foundation;
the Polish State Committee for Scientific Research 
under contract No.2P03B 17017; 
the Ministry of Science and Technology of Russian Federation;
the National Science Council and the Ministry of Education of Taiwan;
the Japan-Taiwan Cooperative Program of the Interchange Association;
and  the U.S. Department of Energy.


\clearpage

\begin{table}
\begin{center}
\caption{The contributions to the systematic error on the branching fraction.}
\label{tbl:syserror}
\begin{tabular}{lr}
\hline
Photon reconstruction                & $\pm5.3\%$ \\
Charged track reconstruction         & $\pm4.7\%$ \\
Charged kaon selection               & $\pm1.8\%$ \\
Charged pion selection               & $\pm1.1\%$ \\
$\KS$ reconstruction                 & $\pm1.2\%$ \\
$\PZ$ reconstruction                 & $\pm3.1\%$ \\
SFW efficiency and $\pi^0/\eta$ veto & $\pm6.8\%$ \\
Best candidate selection             & $\pm2.4\%$ \\
SFW sideband uniformity              & $\pm4.7\%$ \\
Signal MC statistics                 & $\pm3.0\%$ \\
Number of $\BBbar$ events            & $\pm2.3\%$ \\
\hline
Total systematic error               & $\pm12.4\%$ \\
\hline
\end{tabular}
\end{center}
\end{table}

\myfigure{0.70}{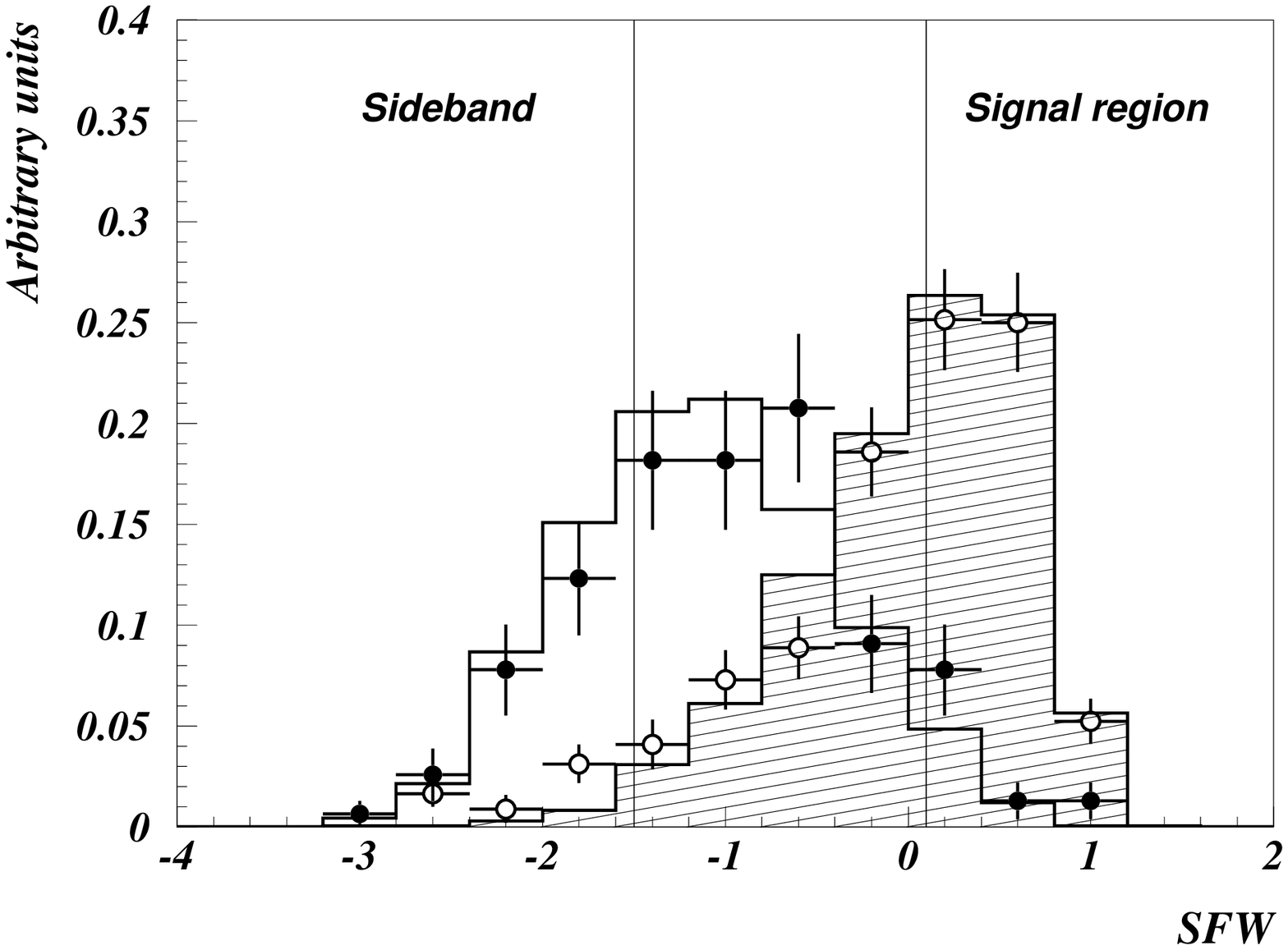}{ The distribution of the SFW variable described in
	the text.  The off-resonance background data (solid circles) and the
	$\qqbar$ MC expectation (open histogram) are compared with the signal
	distribution of $B{\to}D\pi$ data (open circles) and signal MC
	(hatched histogram). }

\myfigure{0.80}{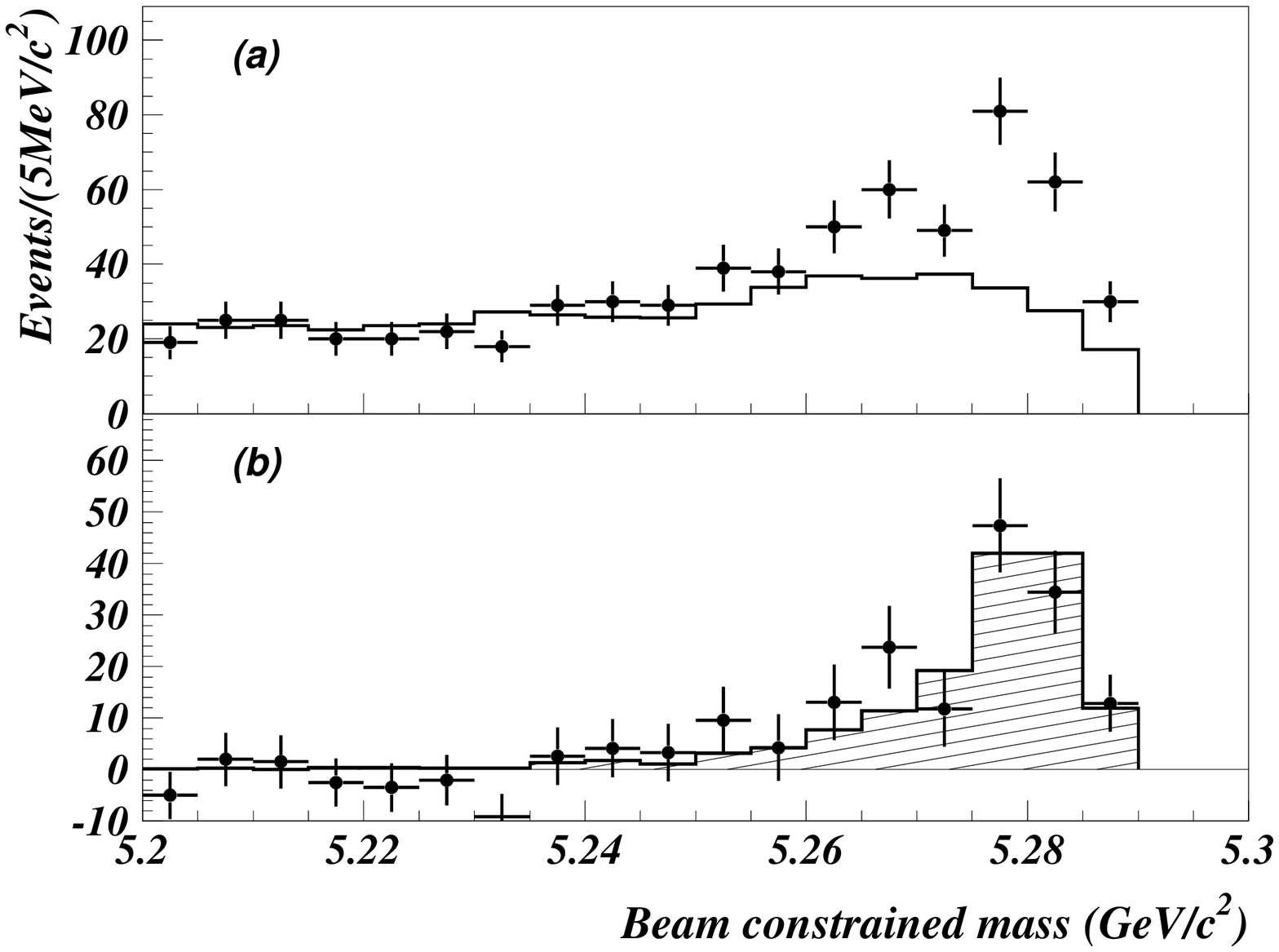}{ The beam constrained mass ($\MB$) distribution (a)
	compared with the total background $\qqbar$ and $\btoc$ (open
	histogram); (b) after background subtraction compared with the
	signal MC expectation (hatched histogram). }

\myfigure{0.5}{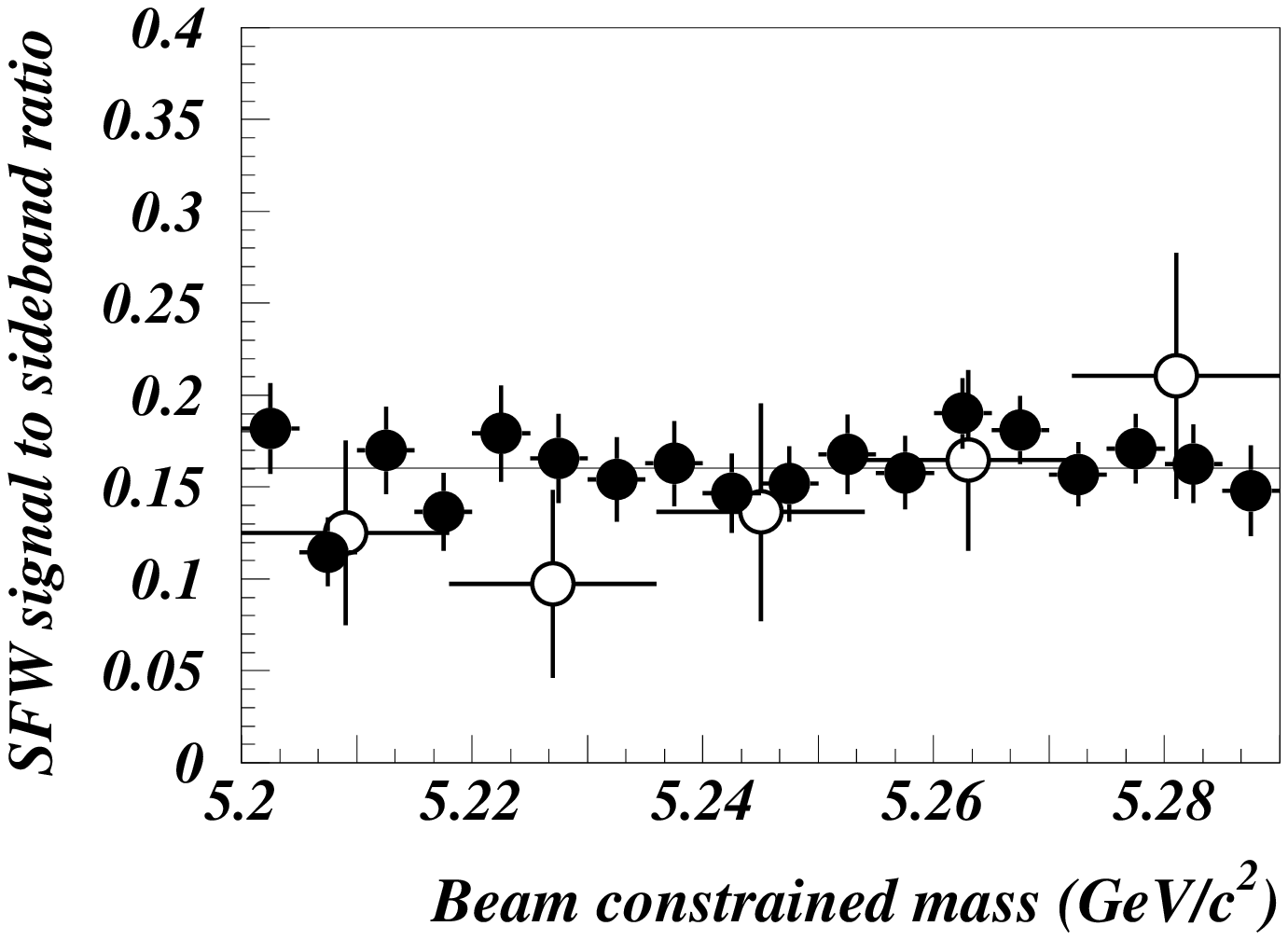}{ The ratio of the background events in the SFW
	    signal region to the sideband region as a function of $\MB$ from
	    the $\qqbar$ MC (solid circles) and off-resonance data (open
	    circles). }

\myfigure{0.7}{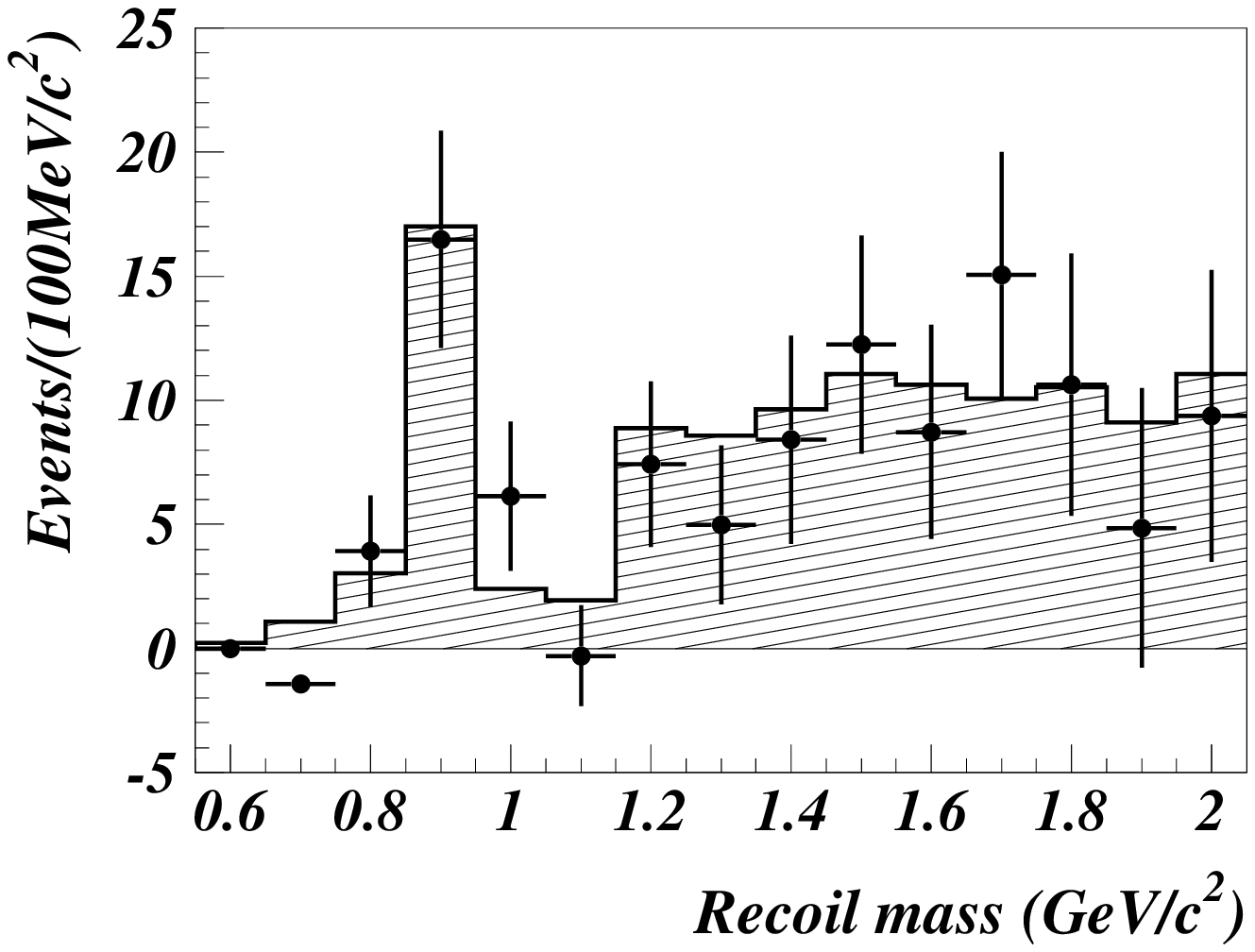}{ The recoil mass ($\MXs$) distribution after
	background subtraction compared with the signal MC expectation
	(hatched histogram).  }

\myfigure{0.7}{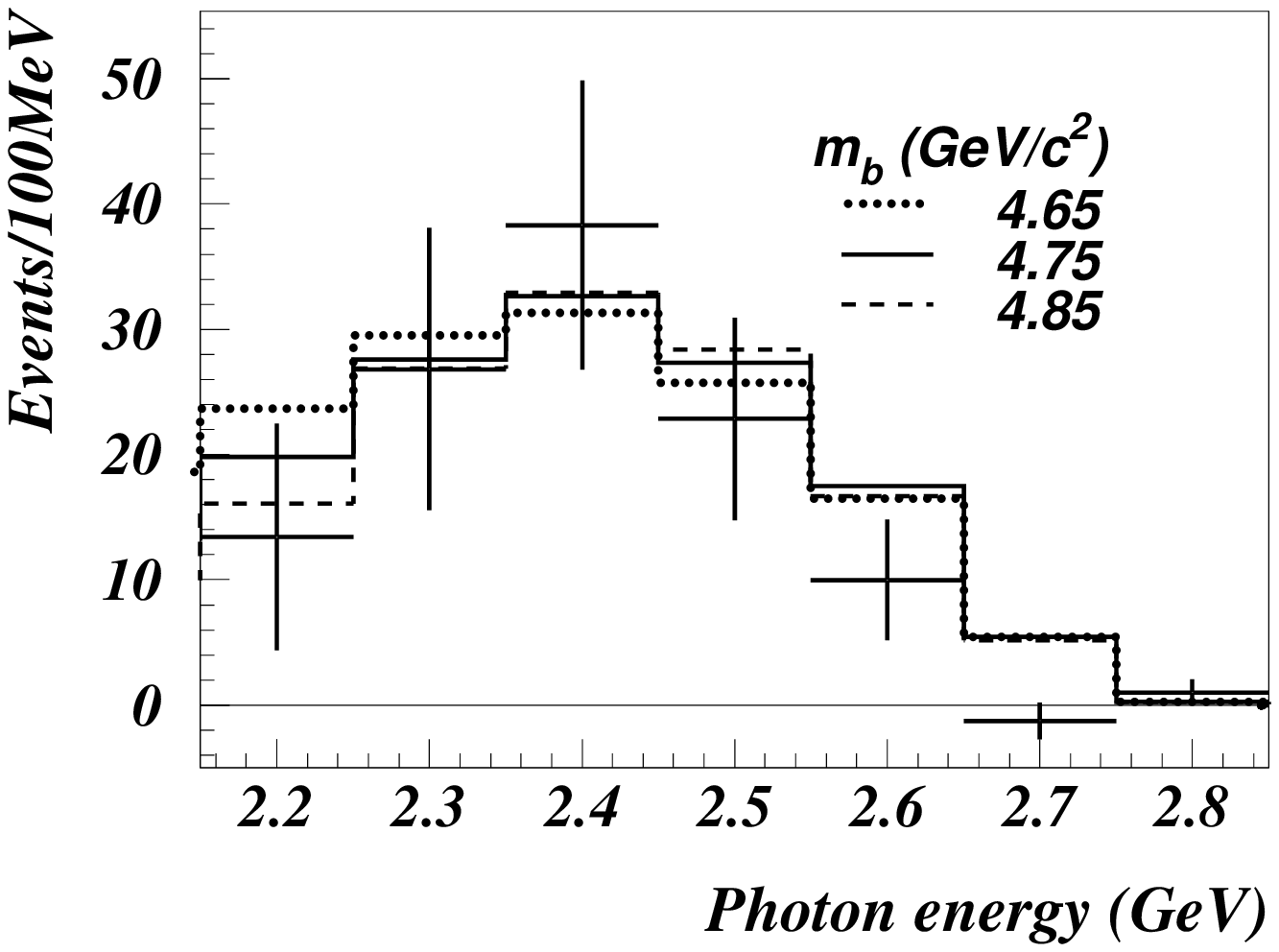}{ The photon energy spectrum, background
	subtracted and corrected for the cut-off on $\MXs$.  The data points
	are compared with signal MC expectations for three different values
	of $m_b$. }


\begin{thebibliography}{99}

\bibitem{bib:xsgam-sm}
K.~Chetyrkin, M.~Misiak, M.~M\"unz,
Phys.~Lett. B 400 (1997) 206;
Erratum ibid. B 425 (1998) 414.

\bibitem{bib:xsgam-nonsm}
For example,
F.~Borzumati, C.~Greub,                              
Phys.~Rev. D 58 (1998) 074004;
M.~Ciuchini, G.~Degrassi, P.~Gambino, G.~F.~Giudice, 
Nucl.~Phys. B 534 (1998) 3;
C.~Bobeth, M.~Misiak, J.~Urban,                      
Nucl.~Phys. B 567 (2000) 153.

\bibitem{bib:kekb}
KEK accelerator group, KEKB B-Factory Design Report, 
KEK Report 95-7 (1995), unpublished.

\bibitem{bib:belle}
Belle Collaboration, K. Abe et al., The Belle Detector, KEK
Progress Report 2000-4 (2000), to be published in Nucl.~Instr.~Meth.~A.

\bibitem{bib:geant}
R.~Brun et al., GEANT 3.21, CERN Report No. DD/EE/84-1 (1987).

\bibitem{bib:mc}
We use the QQ $B$ meson decay event generator
developed by the CLEO collaboration
(http://www.lns.cornell.edu/public/CLEO/soft/QQ), 
and the JETSET program, T. Sj\"ostrand, ``PYTHIA 5.6 and
JETSET 7.3: Physics and manual,'' CERN-TH-6488-92, to hadronize $\qqbar$
and $\Xs$ final states.


\bibitem{bib:hadronB}
Belle Collaboration, K.~Abe et al., hep-ex/0103041, submitted to
Phys. Rev. Lett.

\bibitem{bib:belle-mixing}
Belle Collaboration, K.~Abe et al., Phys.~Rev.~Lett. 86 (2001) 3228.

\bibitem{bib:fox-wolfram}
G. Fox and S.~Wolfram,
Phys.~Rev.~Lett. 41 (1978) 1581.

\bibitem{bib:ushi-thesis}
Y.~Ushiroda, PhD Thesis, Kyoto University (2001), unpublished, see
http://belle.kek.jp/bdocs/theses.html.

\bibitem{bib:kagan-neubert}
A.~L.~Kagan, M.~Neubert,
Eur.~Phys.~J. C 7 (1999) 5.

\bibitem{bib:ali-priv}
The $m_b$ parameter range is decided on the basis of a private
communication with A.~Ali.

\bibitem{bib:lepsldcdf-semilep}
ALEPH, CDF, DELPHI, L3, OPAL and SLD Collaborations, CERN-EP-2000-096,
hep-ex/0009052.

\bibitem{bib:uraltsev}
N.~Uraltsev, Int. J. Mod. Phys. A14 (1999) 4641.

\bibitem{bib:excl-radb}
CLEO Collaboration, T.~Coan et al.,
Phys.~Rev.~Lett. 84 (2000) 5283.

\bibitem{bib:excl-radb-belle}
M.~Nakao (Belle Collaboration), Studies of Radiative $B$ Meson Decays
with Belle, Proceedings of the 30th International Conference on High
Energy Physics, July 2000, Osaka.

\bibitem{bib:cleo-95}
CLEO Collaboration, M.~Alam et al., Phys. Rev. Lett. 74 (1995) 2885.

\bibitem{bib:aleph-98}
ALEPH Collaboration, R.~Barate et al., Phys. Lett. B 429 (1998) 169.

\end{thebibliography}
\end{document}